\documentclass[aps,pre,showpacs,floatfix,superscriptaddress,11pt]{revtex4}
\usepackage{graphics,amsmath,epsfig}
\usepackage{amssymb}
\usepackage{bm}

\def\be{\begin{equation}}
\def\ee{\end{equation}}
\def\bea{\begin{eqnarray}}
\def\eea{\end{eqnarray}}
\def\la{\label}
\def\bsea{\begin{subeqnarray}}
\def\esea{\end{subeqnarray}}

\begin{document}

\setcounter{page}{1}

\title[]{Coarsening Kinetics of a Two Dimensional $O(2)$ Ginzburg-Landau Model: 
             Effect of Reversible Mode Coupling}
%\title[]{Coarsening Dynamics in Two Dimensional XY model with Mode Coupling Terms}

\author{Keekwon \surname{Nam}}
\affiliation{Department of Physics, Changwon National University, Changwon 641-773, Korea}

\author{Bongsoo \surname{Kim}}
\affiliation{Department of Physics, Changwon National University, Changwon 641-773, Korea}

\author{Sung Jong \surname{Lee}}
\affiliation{Department of Physics, University of Suwon, Kyonggi-do 445-743, Korea}

%\affiliation{School of Computational Sciences, Korea Institute for Advanced Study, Seoul 
%130-722, Korea}

\date{\today}
%\date[]{Received April 30, 2008}

\begin{abstract}

We investigate, via numerical simulations, the phase ordering kinetics of  
a two-dimensional soft-spin $O(2)$ Ginzburg-Landau model when a reversible mode 
coupling is included via the conserved conjugate momentum of 
the spin order parameter (the model E). 
Coarsening of the system, when quenched from a disordered state to 
zero temperature, is observed to be enhanced by the existence of the mode coupling terms. 
The growth of the characteristic length scale $L(t)$ exhibits an effective 
super-diffusive growth exponent that can be interpreted as a positive logarithmic-like 
correction to a diffusive growth, i.e., $L(t) \sim (t \ln t)^{1/2} $. 
%We conjecture that, in this model, the mobility of a vortex in a vortex pair diverges 
%logarithmically in the linear size of the vortex pair. Then the length scale is 
%expected to show a diffusive growth multiplied by a positive logarithmic correction 
%as $L(t) \sim (t \ln(t/t_0 ))^{1/2}$. 
In order to understand this behavior, we %the simulation results can be in with 
introduced a simple phenomenological model of coarsening based on the  annihilation
dynamics of a vortex-antivortex pair, incorporating the effect of vortex inertia and
logarithmically divergent mobility of the vortex. With a suitable choice of the parameters,
numerical solutions of the simple model can fit the full simulation results very adequately.   
The effective growth exponent in the early time stage is larger due to the effect of the
vortex inertia, which crosses over into late time stage characterized by 
positive logarithmic correction to a diffusive growth.   
We also investigated the non-equilibrium autocorrelation function from which the 
so called $\lambda$ exponent can be extracted. We get $\lambda \simeq 1.99(2)$ which is 
distinctively larger than the value of $\lambda \simeq 1.17 $ for the  purely dissipative
model-A dynamics of non-conserved $O(2)$ models.

\end{abstract}



\maketitle

%relaxation \sep superconducting arrays \sep frustration
%\pacs{74.50+r, 67.40.Fd}

\section{INTRODUCTION}
\setcounter{equation}{0}

%Usually characteristic topological defects such as vortices or domain walls are generated 
%in the initial disordered state, %depending on the dimension of the order parameter
%and the annihilation of these defects provides the main mechanism of coarsening 
%and phase ordering in the system. The observed self-similarity of these coarsening 
%systems at different time instants, is usually represented by the so-called {\em 
%dynamic scaling hypothesis} of the equal-time spatial correlation function of the order 
%parameter.

The dynamics of statistical systems quenched from high temperature disordered states 
to low temperature ordered states has been a subject of 
interest for several decades 
\cite{gunton_droz,ordering_review, furukawa,binder,bray94_adv_phys,onuki,cugli2010}. 
In typical situations, the average length scale $L(t)$ of the ordered domain grows in time 
as a power law $L(t) \sim t^{1/z}$, where the growth exponent $1/z$ depends on the dimension 
of the space and that of the relevant order parameters, in addition to the conserved or 
non-conserved nature of the latter in the relaxation dynamics \cite{bray94_adv_phys}. 
Usually characteristic topological defects such as vortices or domain walls are generated 
in the initial disordered state, %depending on the dimension of the order parameter
and the annihilation of these defects provides the main mechanism of coarsening 
and phase ordering in the system. The observed self-similarity of these coarsening 
systems at different time instants, is usually represented by the so-called {\em 
dynamic scaling hypothesis} of the equal-time spatial correlation function of the order 
parameter.
Dynamic scaling hypothesis has been an important ingredient in deriving
the properties of phase ordering in the late time stage. For example, in combination 
with the so called energy scaling method \cite{br94}, domain growth laws could be 
extracted for almost all model systems with purely dissipative dynamics.        

Previous works on phase ordering focused mostly on the cases where 
the effect of dissipation dominates the coarsening and the motion of the topological 
defects. On the other hand, in reality, there exist various interesting 
systems exhibiting dynamic processes that cannot be described solely by dissipative 
dynamics. One example can be found in the case of magnetic spin systems, where the spins 
are influenced by neighboring spins via precession interaction terms that are energy 
conserving \cite{ma-mazenko}. These elements in the dynamics are called reversible 
(non-dissipative) mode coupling. 
In fact, among the various model systems (Model $A$, $B$, $C$, $E$, $F$, $G$, $H$, $J$) 
classified by Hohenberg and Halperin\cite{hohen77} that are known to describe the dynamic 
critical phenomena, only models $A$, $B$, and $C$ are based on dissipative dynamics alone, with 
the remaining models $E$, $F$, $G$, $H$, and $J$ retaining reversible mode couplings. 
Hence, it is quite natural to extend  studies on the phase ordering dynamics to these 
model systems. 

Indeed some works have been carried out along this direction.
A most familiar example is the phase separation dynamics of a binary fluid, in which reversible coupling
between the hydrodynamic flow  and the relative concentration, advection of the order parameter field by
the fluid flow, plays an important role \cite{siggia79,puri10}.
Effects of hydrodynamic flow on the phase ordering kinetics and defect dynamics of nematic 
liquid crystals have been studied \cite{denniston01,toth02,svensek02,blanc05}.
Influence of precession on the phase ordering of an isotropic Heisenberg magnet in three dimensions 
(the model J \cite{ma-mazenko}) has also been studied \cite{das-rao2000}. 

A particularly interesting example is the phase ordering kinetics of the Bose gas, i.e., time 
evolution of the Bose-Einstein condensation,  which was  studied via 
the Gross-Pitaevski (GP) equation in two and three dimensions, demonstrating the importance of 
the reversible Josephson precession term in the dynamics \cite{sachdev96a,sachdev96b}.
Alternative approach to this problem is to use the appropriate stochastic model known as the model $F$
\cite{hhs76} in which the complex order parameter field $\psi$ is (both statically and dynamically) 
coupled to the conserved real field $m$.   
 
In this work, we investigate the phase ordering process of systems governed by
a simpler model, namely, the model E \cite{hhs76} in which there is no static coupling between 
$\psi$ and $ m$. Specifically, we focus our investigation on the effect of the reversible
 spin precession term on the phase ordering process in the soft spin $O(2)$ 
models  in two dimensions. We especially try to compare the characteristics of the phase 
ordering dynamics of these model systems with those in the case where dissipative dynamics 
alone is considered.     
The linearized hard spin version of this model was employed by Nelson and 
Fisher \cite{nelson-fisher77} to describe the behavior of the spin wave in the 
two-dimensional anisotropic ferromagnet and the hydrodynamics of the third sound propagation 
in thin film of $He^4$. In \cite{sachdev96a,sachdev96b}, the same set of linearized equations was 
used to demonstrate the crucial role played by the non-dissipative precession term in the phase 
ordering process of the defect free case in the two dimensional XY model. 

In terms of phase ordering dynamics, $O(2)$ models in two dimensions are of particular 
interest because the aforementioned energy scaling method do {\em not} provide a definitive 
result on the domain growth law.  
%Rather, in the case of purely dissipative relaxation dynamics,
%from a phenomenological argument the growing length scale was shown to exhibit a logarithmic 
%correction\cite{korshu_mob,ryskin,rojas} to the diffusive growth as $L(t)\sim (t/\ln t)^{1/2}$. 
In equilibrium, the two dimensional ferromagnetic {\em XY} model exhibits a 
Berezinskii-Kosterlitz-Thouless (BKT) transition at $T_{BKT}$ due to the unbinding of 
vortex-antivortex pairs \cite{berez70}. 
Below $T_{BKT}$, the system has a quasi-ordered phase which is characterized by 
an algebraic decay of the order parameter correlation function for long distances. 
The critical exponent governing this power law decay decreases {\it continuously} down to 
zero temperature. That is, the system is critical at equilibrium for all non-vanishing 
temperatures below $T_{BKT}$. Therefore, the coarsening dynamics in this model  at 
finite temperatures is expected to exhibit {\em critical} dynamic scaling instead of 
simple dynamic scaling.    

Large number of works using the purely dissipative dynamics 
have been carried out on the coarsening dynamics of the two dimensional XY/O(2) models 
\cite{loft87,toyo_honda,mondello,bh,pr,toyo,bp,lm92,yurke93,blundell94,lee_lee_kim,
pbr,mazenko97,bray97,mw,rr,bbj,bray2000,ying2001,jeon2003,qm,koo2006,az,lz}.
After all these efforts, it is now agreed that, in the phase ordering dynamics of ordinary 
{\em $O(2)$} models in two dimensions without mode coupling terms 
and with non-conserved order parameter, the growing length scale exhibits a logarithmic correction 
to the diffusive growth as $L(t)\sim (t/\ln t)^{1/2}$. Here, the logarithmic correction 
can be attributed to a logarithmic divergence (with the system size) of the effective friction
constant (or, inverse mobility) of a moving vortex in the dissipative dynamics with 
non-conserved order parameter \cite{pleiner,ryskin,korshunov,nn,jeli2010}. 

In a related simulation work \cite{jeon2003} on the coarsening dynamics of superconducting Josephson 
junction arrays based on the dynamics of resistively-shunted junction model \cite{bjkim}, it was revealed 
that there is {\em no} logarithmic correction in the growth law, resulting in purely 
diffusive growth law. This 
absence of logarithmic correction was
understood in terms of finiteness of the effective 
friction constant (or inverse mobility) of a moving vortex in the limit of large system size,
which is due to the particular type of dissipative coupling in Josephson junction arrays\cite{jeon2003}. 

In addition, the coarsening dynamics of the two dimensional XY model with a purely 
Hamiltonian dynamics \cite{koo2006,az,leoncini98,cerruti2000,comment} exhibited the growing 
length scale with an apparent late-time power-law growth as $L(t) \sim t^{1/z} $ with 
the exponent $1/z$ { \em larger} than the diffusive exponent $1/2$. 
This kind of (dynamics-dependent) non-universal growth law motivated us to 
investigate further the phase ordering dynamics in related model systems.

Actually, in the model E, we find that the growing length scale $L(t)$ exhibits an apparent 
super-diffusive  growth as $L(t) \sim t^{1/z} $ with the "effective" dynamic exponent $1/z$ a 
little larger than the diffusive exponent $1/2$.  This can be contrasted to the case when reversible
 mode couplings are absent where the apparent growth exponents with values a little smaller than 
$1/2$ are obtained due to the (negative) logarithmic corrections mentioned above.   
The apparent super-diffusive exponent  in the case of reversible mode coupling
may suggest that the asymptotic behavior can be represented as a diffusive growth with 
a positive logarithmic corrections of the form $L(t) \sim (t \ln t)^{1/2}$, 
as discussed in detail below.

One can extract another growing length scale from the excess energy relaxation.
It is expected that the so-called energy scaling method \cite{br94} can be applied to the present 
case as well since the reversible contribution does not dissipate energy. 
This method gives the relation $\Delta E(t) \sim L^{-2}_E (t) \ln  L_E (t)$ (between the excess 
energy $\Delta E(t)$  versus a growing length $L_E (t)$). 
We find that the length scale $L_E (t)$ derived from  the above relation agrees with 
the growing length scale of the domains $L(t)$ in almost the entire time. 

In order to understand the characteristics of the length scale growth in the present system,
we noted that the coarsening and the resultant growth of the length scale are determined  
by the annihilation of vortex-antivortex pairs.    
Therefore, it is plausible to assume that the time $t$ that it takes for the system to
grow up to a length scale $L(t)$ will correspond to the time that it takes for 
vortex-antivortex pairs of separation $L(t)$ to annihilate. Based on this picture,      
we devised a simple phenomenological model of dynamics for the annihilation of  
a vortex-antivortex pair where we incorporate the effect of vortex inertia and 
logarithmically divergent mobility of the vortex \cite{bishop, kamppeter1, kamppeter2}.     
We found that this model can describe closely the simulation results. 
%We find that the inertial effect dominates in the early time stage, while the weak 
%logarithmic effect dominates in the late time stage.     
%It is plausible to assume that these enhancement of coarsening arises from some weak 
%correction (of logarithmic type) due to the existence of a logarithmic divergence in the 
%mobility of an isolated vortex as a function of the typical length scale in the problem. 
%Typical length scale would be determined by the system size for an isolated vortex or the 
%linear size $R$ for a vortex-antivortex pair. 
We  argue that the microscopic mechanism of these coarsening features including the 
growth law, is closely related to the fact that there exists one conserved quantity 
in this model. That is, the $m$ field component (see the next section), which is 
conjugate to the planar spin, is conserved. This corresponds to the third ($z$-) component 
of the spin in Heisenberg spin systems with easy plane anisotropy \cite{nelson-fisher77}. 
Due to the conservation of this component, there appear {\em propagating} spin wave modes 
at low temperatures below $T_{KT}$ where the vortices and antivortices are neglected.
In the case of coarsening dynamics where there exist a lot of vortices and antivortices
in the initial stage, we do not expect that the spin waves are fully propagating. We expect
rather that, due to the interaction of the vortices with the background vortices and 
antivortices, they propagate only within a short range and then scatter off the vortices 
(and antivortices) in a complicated manner such that the mobility of an isolated vortex  
is enhanced with a logarithmic depence on the size of the system.
% (or the size of a vortex-antivortex pair).  

Spin autocorrelation function $A(t)$ also was calculated which is expected to be related 
to the growing length scale $L(t)$ through a new exponent $\lambda$ as
$  A (t) \sim L^{-\lambda } (t) $.    
We could extract the value of $\lambda$ in the long time limit as $\lambda \simeq 1.99(2)$.
This result may be interpreted as $\lambda = d = 2$ for the present case. .
We note that this value is quite distinct from the corresponding value of $\lambda \simeq 1.17 $ 
for the case of no reversible mode coupling which was calculated both theoretically as well as 
numerically \cite{bh,newman90,newman90_2,lm92,lee_lee_kim}.

%\section{The $O(2)$ model with reversible mode coupling terms: the model $E$ }
\section{The  model $E$ }

We consider the ordering kinetics of the soft spin model known as the model $E$ whose 
Hamiltonian is given by  
\begin{equation} 
H[\vec{\phi} , \sigma]  = \int d^2 r \, \Big[ \frac{1}{2} (\nabla \vec{\phi})^2 + 
\frac{1}{4} (\vec{\phi}^2 -1)^2 + \frac{K}{2} m^2 \Big]
\la{eq2.1}
\end{equation}
where $\vec{\phi}$ is the two-component vector order parameter 
$\vec{\phi} =  (\phi_{1}, \phi_{2})$ and $m$ is the third component conjugate to 
spin $\vec{\phi}$ (which corresponds to the $z$-component of the three-dimensional 
magnetization in Heisenberg spin systems with easy plane anisotropy). 
When $K$ is set to zero, the above Hamiltonian becomes equivalent to that of $O(2)$
Ginzburg-Landau model. Therefore, the order parameter has $O(2)$ rotational symmetry 
on $(\phi_1, \phi_2)$-space and the equilibrium average of $m $ is zero. 
The model-$E$ Hamiltonian, (\ref{eq2.1}), does not contain static couplings between 
the order parameter $\vec{\phi}$ and the $m $ field. 
(On the other hand, the Model $F$ does contain such a static coupling.) 
However, as can be seen below, {\em dynamic} couplings between these 
two fields arise from reversible mode coupling terms. The two fields $\vec{\phi}$ and 
$m$ satisfy the following Poisson bracket 
relations \cite{hohen77}
\bea 
\{ \phi_1 (r), m (r^{\prime}) \}  & =  &  -g \phi_2 (r) \delta (r-r^{\prime}), \nonumber \\
\{ \phi_2 (r), m (r^{\prime}) \}  & =  &  g \phi_1 (r) \delta (r-r^{\prime}), \nonumber  \\
\{ \phi_1 (r), \phi_2 (r^{\prime} ) \} & =  &  0 
\la{eq2.2}
\eea
where $g$ denotes the strength of the mode coupling that is an analog of the gyromagnetic 
ratio of the spins. These Poisson bracket relations generate the reversible mode 
couplings in the equations of motion, which cause spin precession. They play a key role 
in the critical dynamics \cite{hohen77}. Note that the order parameter $\vec{\phi}$ is not 
a conserved quantity. Instead, conserved quantity in this system is the $m$ component 
which causes the precession of the order parameter $\vec{\phi}$ in the $x-y$ plane of 
the spin space \cite{hohen77}.

The dynamics of the system described by the fields
$\psi_i 's$ ($ = \phi_1, \phi_2, m$) can be written as 
\begin{equation}
{\partial \psi_i \over \partial t}=\sum_{j} \left [  \{ \psi_i, \psi_j \} {\delta H \over
    \delta \psi_j } - \Gamma_{ij} { \delta H \over \delta \psi_j } \right ] 
     + \eta_i (\vec{r},t)
\label{eq2.3}
\end{equation}
where $\Gamma_{ij}$ denotes generalized kinetic coefficients ($\Gamma$ and $D$ below) and 
$ \xi_i$ denotes the thermal noises ( $\eta_1$, $\eta_2$ and $\zeta$ below). 
$\{ A, B \}$ is the Poisson bracket of two generic dynamic variables $A$ and $B$ as 
defined above (\ref{eq2.2}). 
With (\ref{eq2.2}) and the above relation, the equations of motion becomes 
\bea
{\partial \phi_1 \over \partial t} & = & - g \phi_2 { \delta H \over \delta m} \
- \Gamma { \delta H \over \delta \phi_1 }  + \eta_1 (\vec{r},t), \nonumber \\
{\partial \phi_2 \over \partial t} & = &  g \phi_1 { \delta H \over \delta m} \
- \Gamma { \delta H \over \delta \phi_2 }  + \eta_2 (\vec{r},t), \nonumber  \\
{\partial m \over \partial t} & = &  g \Big(-\phi_1 { \delta H \over \delta \phi_2 } \
+ \phi_2 {\delta H \over \delta \phi_1} \Big) + D \nabla^2 { \delta H \over \delta m } \
 + \zeta (\vec{r},t).  
\la{eq2.4}
\eea
where $\Gamma$ and $D$ are the kinetic coefficients, and the Gaussian thermal noises
$\vec{\eta}$ and $\zeta$ at temperature $T$ satisfy 
\bea
\langle \eta_{i}(\vec{r}, t) \eta_{j} (\vec{r^{\prime}}, t^{\prime}) \rangle & = & 2
\Gamma T  \delta_{ij} \delta (\vec{r}-\vec{r^{\prime}})\delta (t-t^{\prime}), \nonumber \\
\langle \zeta(\vec{r},t) \zeta (\vec{r^{\prime}},t^{\prime}) \rangle & = & - 2 D T  
 \nabla^2 \delta(\vec{r}-\vec{r^{\prime}}) \delta (t-t^{\prime}), \nonumber \\
\langle \eta_i(\vec{r},t) \zeta (\vec{r^{\prime}},t^{\prime}) \rangle & = & 0
\la{eq2.5} 
\eea

The equation of motion (\ref{eq2.3}) can be explicitly written as 
\bea
{\partial \phi_1 \over \partial t} & = & -g K \phi_2 m 
+ \Gamma \big[  \nabla^2\phi_1 + (1- \phi_1^2 -\phi_2^2 ) \phi_1 \big] + \eta_1 (\vec{r},t), \nonumber  \\
{\partial \phi_2 \over \partial t} & = & g K \phi_1 m
+ \Gamma \big[ \nabla^2 \phi_2 + (1-\phi_1^2 -\phi_2^2 ) \phi_2 \big] + \eta_2 (\vec{r},t), \nonumber  \\
{\partial m \over \partial t} & = &  g \vec{\nabla} \cdot \big( \phi_1 \nabla \phi_2 \
 - \phi_2 \nabla \phi_1 \big) + D K  \nabla^2 m  + \zeta (\vec{r},t).  
\la{eq2.6}
\eea
We can see again that, in the limit of $g=0$, the order parameter and $m$ are decoupled, 
resulting in the equation of motion for the order parameter equivalent to that of the time-dependent
Ginzburg-Landau equation for $O(2)$ model.  
The equation for $m$ in (\ref{eq2.6}) can be written in the form of a continuity equation 
\bea
{\partial m \over \partial t} & = & -   \nabla \cdot  \vec{J}_{m}, \nonumber \\
\vec{J}_{m} & \equiv &   g ( \phi_2 \nabla \phi_1 - \phi_1 \nabla \phi_2 )  \
   - D K  \nabla m + \vec{\xi} (\vec{r},t)    
\la{eq2.6a}
\eea
where $\vec{J}_{m} $ is the corresponding current density.
We see from (\ref{eq2.6a}) that $m$ is a conserved quantity. In (\ref{eq2.6a}), 
the new thermal noise $\xi$ satisfies
\bea
\zeta (\vec{r},t)  & \equiv &  -\nabla \cdot \vec{\xi}, \nonumber   \\ 
\langle \xi_i(\vec{r}, t) \xi_j (\vec{r^{\prime}}, t^{\prime}) \rangle & = & 2
D T  \delta_{ij}\delta (\vec{r}-\vec{r^{\prime}})\delta (t-t^{\prime})
\la{eq2.7}
\eea
Note that the equations of motion (\ref{eq2.6}) are invariant under the transformation 
$m \rightarrow -m$, $ \phi_1  \rightarrow \phi_2 $, and
$ \phi_2 \rightarrow \phi_1 $. 
%Due to this characteristics, the model is also called the symmetric planar ferromagnet.        

An important physical element brought on by the reversible mode coupling term is the 
existence of the propagating spin wave especially when we can neglect the existence of 
vortices and antivortices. In reference \cite{sachdev96a}, in the quench dynamics of 
Gross-Pitaevskii equation, a linear growth of $z=1$ was found, which was conjectured 
to be due to the existence of propagating spin waves. Here in the case of model $E$, 
however, we find only a minor enhancement with logarithmic correction to a diffusive 
growth. This implies that the spin wave here is not
fully propagating but interacting in some complicated manner with the vortices or antivortices,
resulting in suppression of the propagation due to scattering with the vortices and 
antivortices. See also the discussions in section { \bf IV} on related questions.

%As will be seen in the simulation results, the propagating mode tends to enhance the 
%motion of vortices and the annihilation thereof in comparison with the case of damped 
%spin wave, leading to considerable change in the growth law of the phase ordering kinetics.   

\section{Simulation Results and Discussions}

In order to investigate the phase ordering kinetics of the model $E$, 
we integrated the spatially discretized form of the equations of motion (\ref{eq2.5}) 
at zero temperature $T=0$, i.e., in the absence of thermal noise, starting from a
random initial configuration (corresponding to an equilibrium at $T= \infty$). 
As for the values of the parameters, we put $\Gamma =D=K= 1$. 
In this work, we fix the intensity of the precession $g=1$.
Simulations are performed on systems with spatial discretization of square lattice
type of dimensions up to $3000 \times 3000$ with periodic boundary condition.
For the integration time interval of Euler method, we chose $\delta t = 0.05$. 
The simulation results were the same for smaller time intervals. 

Two main quantities of interest in phase ordering kinetics of this system
are the equal-time spatial correlation of the order parameter defined by
\begin{equation}
C (r,t)  =  {1 \over N} \left < \sum_{i} \vec{\phi}_{i}(t) \cdot \vec{\phi}_{i+r}(t) \right >.
\label{eq3.1}
\end{equation}
%where $< \cdots >$ denotes average over random initial configurations.
and the spin autocorrelation function $A(t)$ defined as 
\begin{equation}
A(t)  =  {1 \over N} \left < \sum_{i} \vec{\phi}_{i}(0) \cdot \vec{\phi}_{i}(t) \right >.
\label{eq3.2}
\end{equation}

The equal-time spatial spin correlation function satisfies the simple dynamic 
scaling (Fig.~1)
\begin{equation}
  C (r,t)   = G(r/ L (t))    
  \la{eq3.3}
\end{equation}
with the correlation length $L (t)$  defined from  $ C (r,t) ]_{r=L (t)}  = C_0$.
Here $C_0$ was conveniently taken as $C_0=0.4$.
The correlation length $L (t) $ is found to grow in time as an apparent power law
\begin{equation}
L (t) \sim    t^{\phi}, \;\;\; \phi \simeq 0.54(1)  
\la{eq3.4} 
\end{equation}

It is interesting to find that the apparent domain growth exponent in the late time regime 
($\phi \simeq 0.54$) is somewhat larger than the diffusive value of $1/2$ (Fig.~2a). 
This can be contrasted to the conventional purely diffusive
 case ($g=0$) where the effective growth exponents obtained numerically are invariably smaller than
$1/2$.      
%It is plausible to assume that these enhancement of coarsening arises from some weak 
%correction (of logarithmic type) due to the existence of a logarithmic divergence in the 
%mobility of an isolated vortex as a function of the typical length scale in the problem. 
%Typical length scale would be determined by the system size for an isolated vortex or the 
%linear size $R$ for a vortex-antivortex pair. 
Now, in order to understand this behavior, we construct a simple phenomenological model of
coarsening of the system as follows. To begin with, we note that the coarsening of the
system is dominated by the annihilation of vortex-antivortex pairs. It is easy to see that, in order
for the system to grow up to a length scale $L$, vortex pairs of sizes on the order of 
$L$ must be already annihilated. Even though this annihilation process of vortex pairs
is very complicated many-body process involving many vortex pairs, we assume that we can
simplify the whole coarsening process (corresponding to the growth of the length scale 
up to $L$) by the annihilation of {\em single} vortex-antivortex pair of size $L$ with 
suitably defined interaction potential and other dynamical parameters. 

From previous works on the dynamics of a vortex in the related model systems of anisotropic 
Heisenberg spin systems (which is based on a collective variable 
approach) \cite{bishop, kamppeter1, kamppeter2}, we might assume that the vortex acts like 
a small particle with finite 'mass' moving under the influence of an external force with 
some (length-scale dependent) mobility. Then the distance $R$ between 
a vortex and an anti-vortex would be described by the following equation of motion 
\begin{equation}
M(R) {{d^2 R} \over {dt^2} } +  {1 \over \mu (R)} {{dR} \over {dt} } =
F(R) =  - { k \over R }.   
\la{eq3.5}
\end{equation}
Here we denote the length-dependent effective mass of a vortex as $ M(R)$ and
the length-dependent mobility of a vortex as $\mu(R)$. We also assume naturally that 
the vortex-antivortex pair is interacting via a Coulombic force $F(R) =-k/R$ in two 
dimensions with a proportionality constant $k$. We will set $k=1$. 

One important remaining question is how we set the functional form of the $M(R)$ and 
$\mu (R)$. From the apparent superdiffusive behavior of the growing length scale obtained
from the simple power law fit with an exponent a little larger than $1/2$, we guessed 
that the growing length scale might be represented, at least in the late time regime,
as $L(t) \sim (t \ln t)^{1/2}$. 
%This implies that the mobility $\mu(R)$ should have a logarithmic dependence on $R$. 
If we neglect the mass term, this kind of positive logarithmic correction can be 
obtained by assuming a logarithmically divergent vortex mobility $\mu(R)$. 
Similar phenomenon of logarithmically divergent mobility was also found in the 
quasi-two-dimensional diffusion of colloids \cite{sane_2009} or diffusion of protein 
molecules on the membranes \cite{saffman_1975} under hydrodynamic 
effect, and also in the vortex diffusion in the anisotropic Heisenberg system in two dimensions
with spin precession \cite{bishop, kamppeter1, kamppeter2}.     
As for the form of $M(R)$, we turned to the anisotropic Heisenberg model where 
the effective mass of a vortex with logarithmic dependence on the length scale was derived 
\cite{bishop, kamppeter1, kamppeter2}.

Motivated by these phenomena in analogous systems, we assumed that the effective mass and 
the mobility are logarithmically dependent on $R$ as
\begin{equation}
M(R) = m_0 + m_1 \ln (R/r_0), \qquad  \mu (R) = \mu_0 + \mu_1 \ln (R/r_0) 
\la{eq3.6}
\end{equation}
where $m_0$, $m_1$, $\mu_0$, $\mu_1$ are constants and $r_0$ denotes a
shortest cutoff length scale in the system (corresponding to the vortex core size).
Note that $m_0$ corresponds to the effective mass at the shortest cutoff length 
scale and similarly for $\mu_0$ for the vortex mobility.   

The way we obtain the growth law from this simple model of vortex-antivortex pair dynamics
is as follows. We first begin with a finite value of the size of vortex-antivortex pair
$R = R^*$. For convenience, we also set the initial velocity of the vortex to be
zero. Then we numerically solve the model equation of vortex-antivortex pair dynamics, (13) and 
(14),  to  obtain the time $\tau$  when 
the size of the vortex-antivortex pair $R$ becomes equal to $r_0$ the lower cutoff length
scale. By plotting the resulting relation $R^*$ and $\tau$ we get the growth law 
of the coarsening dynamics.  

It would be ideal if we could derive the parameters $m_0$, $m_1$, $\mu_0$, and $\mu_1$
from the full dynamic equations, (6) in the absence of thermal noise. 
But we do not know how to proceed 
from this equation to calculate these parameters. We simply tried to fit our simulation 
results (especially the dependence of the growing length scale on time) with the 
numerical solutions of our simple dynamic  model with our tuned values of the 
parameters such that the two results agree most favorably.    

In order to compare the numerical solutions of the simple dynamic model with the simulation 
results, we plot in Fig.~2b $L^2(t)/ t $ versus $t$ from which we can check the behavior of 
the (multiplicative) correction to the diffusive growth. It shows that there exists a smooth 
crossover from early time ($t < 300.0$) behavior with larger slope to later time behavior 
with smaller (slower) slope. The solid line in Fig.~2b represent the results of numerical 
solutions from the simple model of vortex annihilation where we chose the parameters 
$m_0 = m_1 = 1.0 $,  $\mu_0 = 1.0$, and $\mu_1 =0.225$. We see that the numerical solution
exhibits a reasonable agreement with the full simulation results (open circles).   
From our numerical solutions to the simple model we see that the early time behavior comes 
from the effect of the inertial term. If the mobility of the vortex is assumed to be a 
constant independent of the size of the vortex-antivortex pair, then the inertial effect ceases to 
be effective in longer time regime and the correction term would approach a constant plateau. 
However, the real simulations of the coarsening exhibits a steady increase in the correction 
in the late time regime. That is why we incorporated a length-dependent mobility of the vortex 
with a logarithmic dependence on the size of the vortex-antivortex pair. We found that, as 
the coefficient $\mu_1$ of the logarithmic term increases, the later time correction with 
logarithmic behavior gets stronger (data not shown here).       

In the long time limit with large separation between vortices and antivortices, we can 
expect that the inertial effect would be negligible. Therefore, due to the
logarithmic divergence of the vortex mobility, we can analytically get the dominant asymptotic growth 
law as      
\begin{equation}
L(t) \sim (t \ln t)^{1/2} . 
\la{eq3.7}
\end{equation} 
The numerical solution of (13) in the long time region, shown in Fig.~2b confirms this growth law.

In addition, Shown in Fig.~3 are the relaxations of the vortex number density $\rho (t)$ and 
the excess energy density $ \Delta E (t)$ in a log-log plot. 
From the logarithmic slopes, we get  apparent power-law relaxation in time
$\rho (t) \sim    t^{-1.03(1)}$ and $ \Delta E (t) \sim t^{-0.95(1)}$.
In order to understand the relationship between these relaxation behaviors 
and also this apparent growth exponent (\ref{eq3.4}), we note 
that there exists the so called energy scaling relation between the excess energy 
$\Delta E (t)$ and the length scale $L (t)$ of the domain growth for the case of 
$O(2)$ models in two dimensions, i.e., $\Delta E (t) \sim  L^{-2} (t) \ln (L (t)/a_0 )$
with $a_0$ denoting a short distance cutoff corresponding approximately to the 
size of a vortex core. This was first derived by Bray and Rutenberg \cite{br94}. 
Since the reversible mode coupling terms satisfy the energy conservation, this relation 
should be valid in the presence of mode coupling as well.
From the excess energy relaxation (obtained numerically) we can extract a length scale 
$L_{E}(t)$ through the relation 
$\Delta E (t) \sim  L_{E}^{-2} (t) \ln (L_{E} (t)/ a_0 )$. We can also extract another 
length scale $L_V(t)$  from the defect density relaxation that corresponds to the 
average separation between neighboring defects in such a way as 
$\rho (t) \sim L_{V}^{-2} (t)$ which is based on the assumption that the vortices are 
uniformly (and randomly) distributed in two dimensional space.  
These two length scales as well as the numerical sulution to the aforementioned simple model of 
vortex-antivortex aniihilation, together with $L(t)$ are displayed in Fig.~4a.
%%%%%%%%%%%%%%%%%%%%%%%%%%%%%%%%%%%%%%%%

This figure shows that the length scale $L_E (t)$ agrees with the growing length scale of the
domains $L(t)$ in almost entire times, which implies that the above Bray-Rutenberg relation 
is valid in the zero-temperature coarsening of the model E.
 The Bray-Rutenberg relation, together with the growth law (\ref{eq3.7}), 
leads to the asymptotic relaxation behavior for $\Delta E(t)$ as 
\bea
%\rho(t) & \sim & L^{-2}_V (t) \sim L^{-2}(t) \sim \big( t \ln t  \big)^{-1}, \nonumber \\
\Delta E(t) & \sim & t^{-1} \Big(  1+ \frac{\ln (\ln t)}{\ln t}  \Big)
\la{eq3.8}
\eea
Figure~4a also shows that there is considerable difference between the length scale 
$L_V(t) \sim \rho^{-1/2}(t)$ and the length scale $L(t)$. Although this discrepancy appears 
to become smaller in the longer times, their agreement in the long time limit does not seem 
to be guaranteed.  This feature seems to indicate that the assumption of vortices being 
uniformly distributed  is not valid. This is also shown in Fig.~4b depicting the vortex number
density $\rho(t)$ versus $L(t)$ which can be fit for a wide range of time with 
$\rho(t) \sim L(t)^{-1.92(1)}$. 

The nonequilibrium spin autocorrelation function $A (t)$ is expected to be related 
to the growing length scale $L(t)$ through a new non-equilibrium exponent $\lambda$ as  
\begin{equation}
A (t) \sim L^{-\lambda } (t).    
\la{eq3.9}
\end{equation}
We could extract the value of $\lambda$ by plotting $A(t)$ versus $L(t)$ as shown in
the Fig.~5, where we can see that, in the long time limit the value 
of $\lambda$ approaches $\lambda \simeq 1.99(2)$.  This value is much larger than the 
value of $\lambda \simeq 1.17 $ for the case of non-conserved $O(2)$ model with 
no reversible mode coupling \cite{newman90,bh,newman90_2,lm92,lee_lee_kim}.
It appears that the higher mobility of the vortices in the present model causes a 
faster decay of the autocorrelation and hence larger value of the $\lambda$ exponent 
is ensued compared to the case of $O(2)$ model without mode coupling terms.
It would be interesting to prove the conjecture $\lambda = d = 2$ analytically.  

\section{Summary and outlook}

In this work, we studied the zero-temperature coarsening kinetics of the model E in which 
the order parameter field $\psi$ is dynamically coupled with additional conserved field $m$.
We found that the phase ordering kinetics of two dimensional  
$O(2)$ model is modified considerably due to  these reversible mode coupling terms. 
 We can summarize the simulation results as follows. 
The growth of typical length scale in the model E is found to exhibit an apparently 
super-diffusive behavior. 
We introduced a simple phenomenological model for the annihilation dynamics of 
a vortex-antivortex pair incorporating vortex inertia and logarithmically 
divergent mobility of the vortex. This model was shown to describe closely the 
positive logarithmic corrections of the simulation data, where the inertial effect 
dominates in the early time stage, while the positive logarithmic effect of the 
vortex mobility dominates in the late time stage.     

We also investigated the autocorrelation function of the order parameter field. 
The numerical result on the value of $\lambda$ exponent for the present case of the model $E$ 
is approximately $\lambda \simeq 1.99(2)$ 
which is quite distinct from the value $\lambda \simeq 1.17$ for the case of purely 
dissipative dynamics of $O(2)$ model without reversible mode couplings. 
Even though we do not have an analytic proof, it also leads us to 
conjecture that $\lambda = d = 2 $ for the model $E$ which seems to be closely
related to the higher mobility of the vortices as compared to the case of no
mode coupling.

We think that the interaction of propagating spin waves (generated by the mode coupling) 
with the vortices (and antivortices) influence the motion of vortices (and antivortices) 
in such a way that the vortex mobility increases logarithmically on the sizes of 
vortex-antivortex pairs, thereby facilitating the annihilation of the vortex-antivortex 
pairs in the late time stage of coarsening process with a logarithmic correction to a diffusive
growth. Further investigation would be necessary to understand in more detail the interaction 
between the vortex defects as well as that between the vortices and the propagating spin waves.     

In this work, we only considered the phase ordering kinetics of the model $E$ in two 
dimensions, in which static coupling between the order parameter field and the conserved 
field $m$ is ignored. In view of the recent surge of interest in the phase ordering 
kinetics in the ultra-cold atomic gases, it would be worthwhile to extend our study to 
the coarsening of the model $F$ which possess a static coupling between the order parameter 
field and the conserved third component.  In Ref. \cite{sachdev96a}, the superfluid 
ordering kinetics of the Bose gas followed by an instantaneous quench 
from high-temperature normal state was studied using the Gross-Pitaevskii
equation for the order parameter field in both two and three dimensions.  
The order-parameter correlation function was shown to obey a critical dynamic scaling 
for $d=2$ and a simple scaling for $d=3$ with the growing length scale exhibiting a 
linear growth in time, i.e., $L(t) \sim t^{1/z}$ with the dynamic exponent $z \simeq 1.0$ 
due to the presence of propagating spin waves in both dimensions.
Therefore  it would be interesting to carry out a detailed study on the coarsening kinetics of 
the corresponding stochastic model, the model $F$,  in order to see whether the observed 
linear growth in the Gross-Pitaevskii equation is preserved in the model $F$ or not.

Recently, the Bose-Einstein condensate with nonzero spin, i.e., the spinor condensates have 
been experimentally realized for alkali atoms such as $^{23}$Na \cite{stenger} and 
$^{87}$Rb \cite{sadler}.  Interplay between superfluid and magnetic ordering in these 
spinor condensates lead to unusual phases and the corresponding topological defects \cite{ueda}.
One remarkable example is the existence of the nematic superfluid phase and the 
associated half-vortices in the so-called polar phase (observed in sodium atoms), 
which are usually observed in  nematic phase of liquid crystals \cite{moore2006,lamacraft2010}. 
Quite recently, the model $F$ dynamics was generalized and applied to the magnetic domain growth
in the ferromagnetic phase of spinor condensates (observed in rubidium atoms), with and without 
the constraint of the total magnetization conservation \cite{moore2007,lamacraft2007}.  
Extending our study along these fascinating developments in the nonequilibrium dynamics of 
ultracold atoms would be  quite rewarding.

\bigskip

\begin{acknowledgments}
This work was supported by the National Research Foundation of Korea (NRF) grant funded by 
the Korea government (MEST) (No. 2009-0090085). We also thank Korea Institute for Advanced Study 
for providing computing resources on the Linux Cluster System (KIAS Center for Advanced Computation) 
for this work.
%We also acknowledge the support of PLSI program for allowing access to the Grid computing 
%facilities.

\end{acknowledgments}

%\newpage

\begin{figure}[t]
\includegraphics[angle=0,width=8cm]{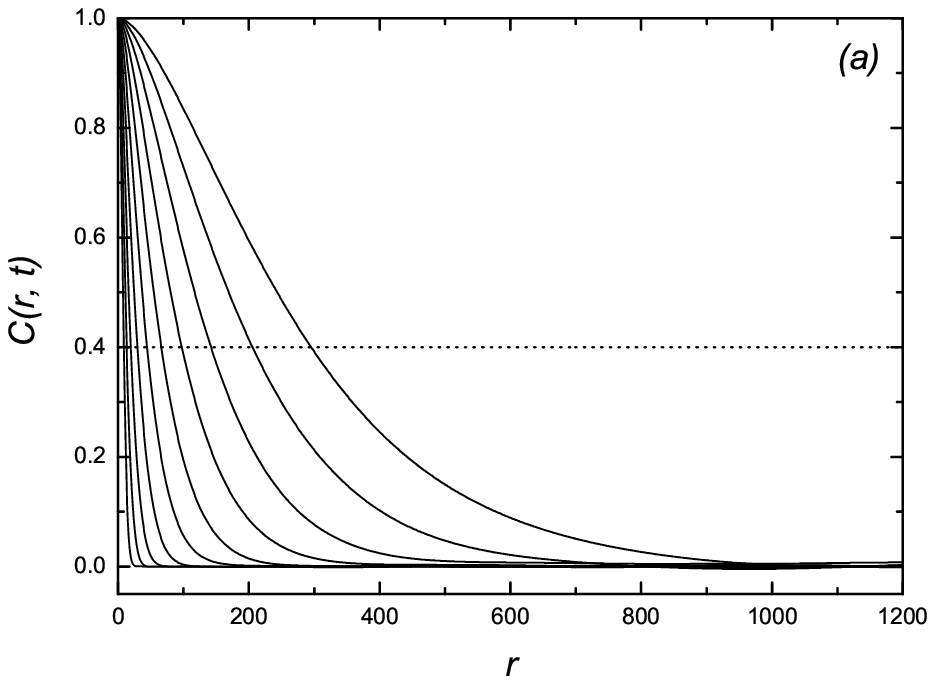}
\includegraphics[angle=0,width=8cm]{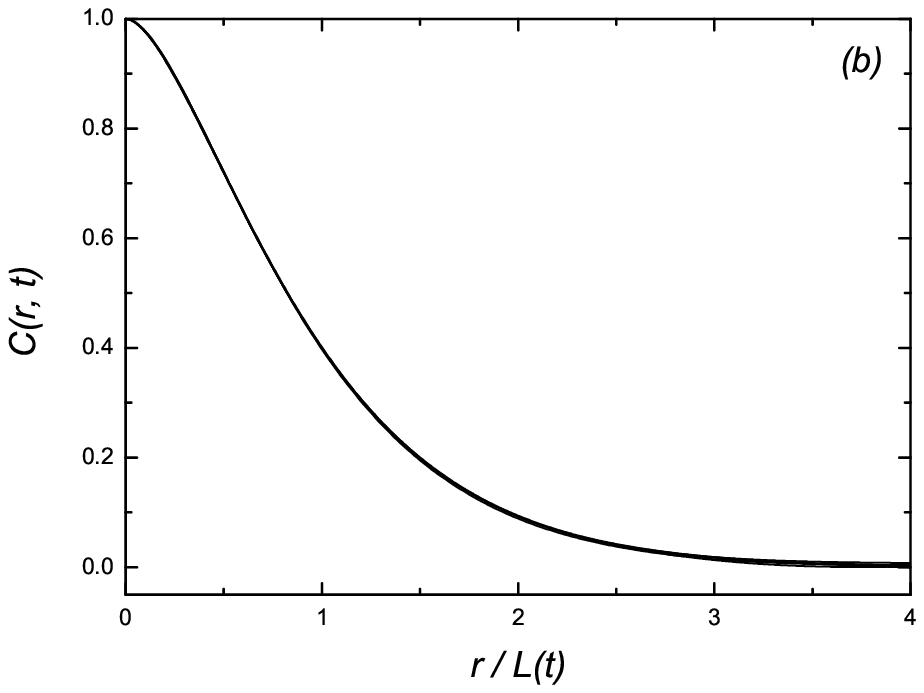}
\caption{ 
(a) The equal time spatial correlation functions for the $O(2)$ order 
parameter of the model $E$ at various time stages ($t=10$, $20$,$40$,$80$,
$160$,$320$,$640$,$1280$,$2560$,$5120$) with the system size 
$3000 \times 3000$ and (b) the scaling collapse of the data in (a)
with the appropriate scaling length $L(t)$ for $t=320$,$640$,$1280$,$2560$, and $5120$ 
as explained in the main text.} \label{corr_scaling}
\end{figure}

\begin{figure}[t]
\includegraphics[angle=0,width=8cm]{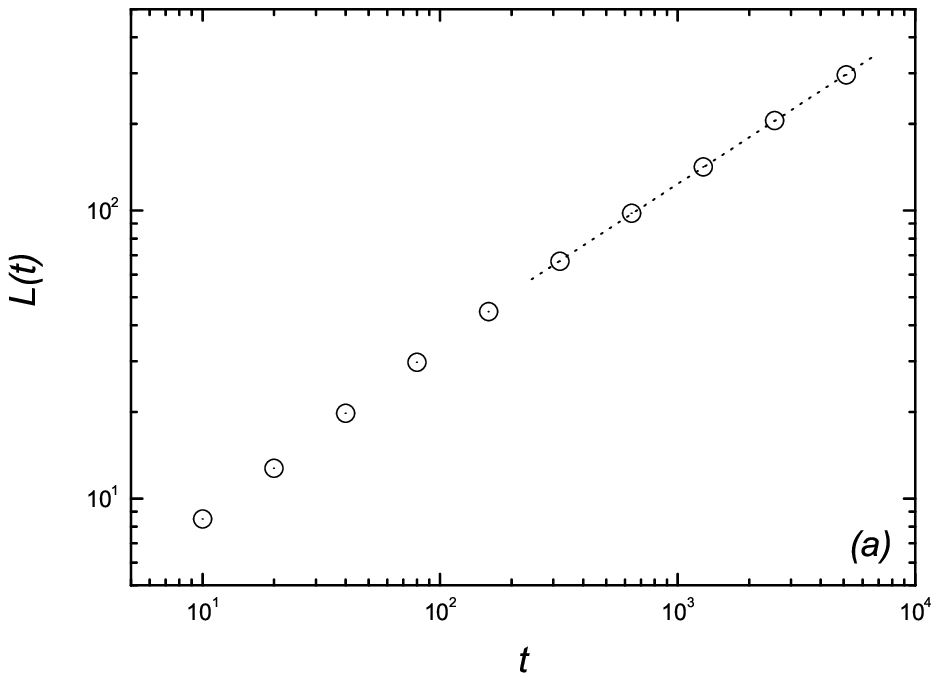}
\includegraphics[angle=0,width=8cm]{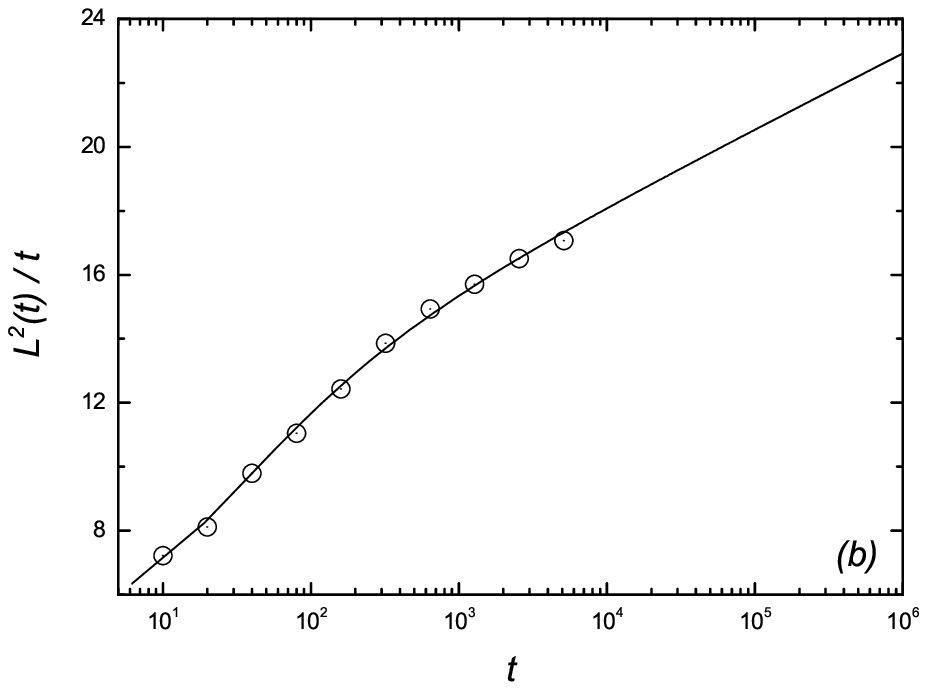}
\caption{ 
(a) Growing length $L(t)$ vs. $t$ (open circles) manifesting an apparent power law 
    growth $L(t) \sim t^{\phi}$ with the effective growth exponent of $\phi \simeq 0.54(1)$
    (dotted line).
(b) ${L(t)^2 }/ t $ vs. $t$ from the full simulation (open circles) is displayed 
   here together with the result of a numerical solution (solid line) of simple 
   dynamic model of vortex-antivortex pair annihilation with a suitably chosen 
   set of parameters that agrees reasonably well with the simulation results.
   Note that the result from the simple model is scaled by a constant factor 
   to fit the full simulation results. 
%Error bars are of the order of the 
%size of the symmbols or smaller. Lines are only guides to the eyes.
} \label{growth_L}
\end{figure}

\begin{figure}[t]
\includegraphics[angle=0,width=8cm]{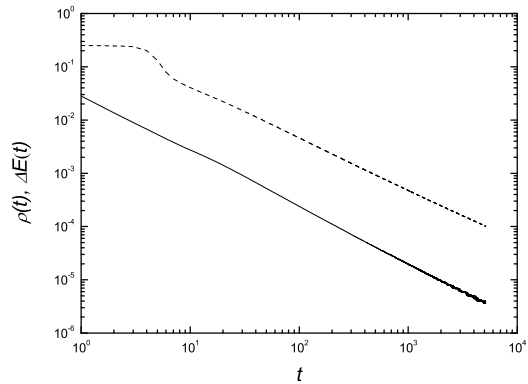}
\caption{ 
Relaxation of the vortex number density $\rho (t)$ (solid line) and
the excess energy $\Delta E (t)$ (dashed line) exhibiting $\rho (t) \sim t^{-1.03(1)}$ and 
$\Delta E (t) \sim t^{-0.95(1)}$ respectively.
} \label{vort_density}
\end{figure}

\begin{figure}[t]
\includegraphics[angle=0,width=8cm]{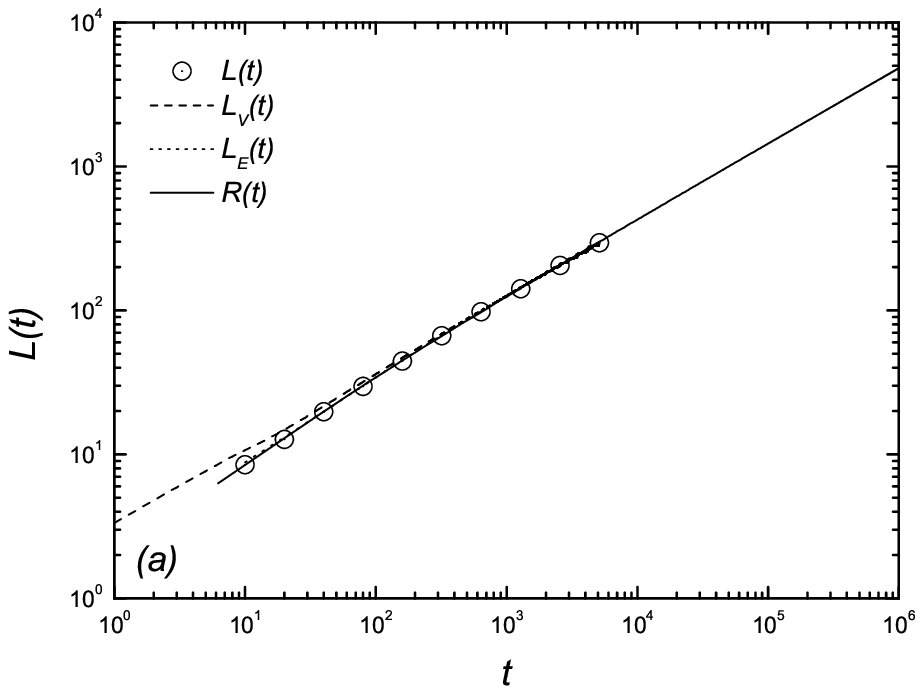}
\includegraphics[angle=0,width=8cm]{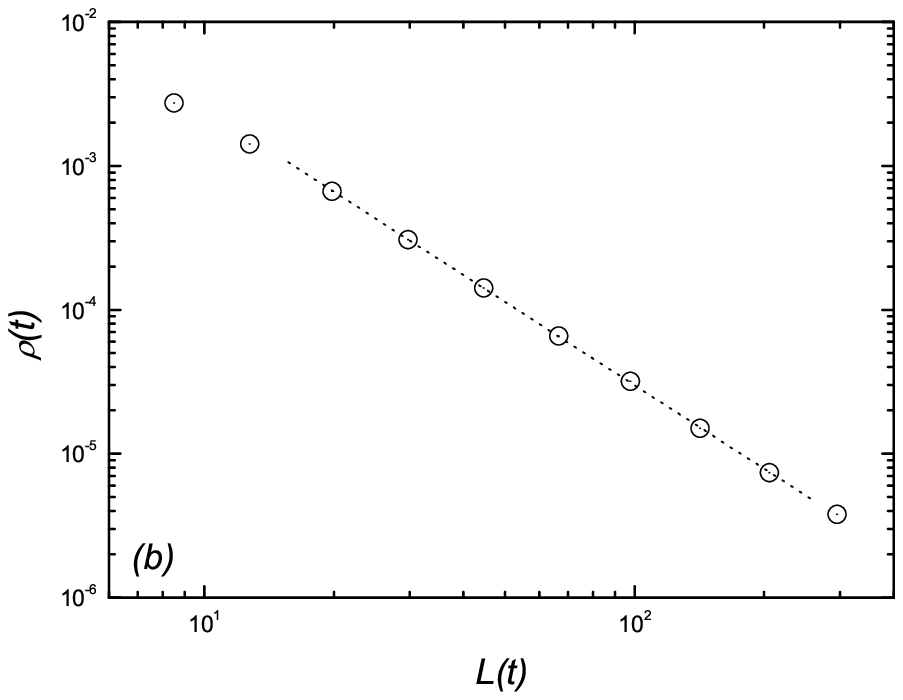}
\caption{ 
 (a) The growing length $L(t)$ (open circles) is shown together with the two length scales 
$L_E (t)$ (dotted line), $L_V (t)$ (dashed line) derived from the energy scaling relation 
and  the decay of the vortex number density, respectively, in addition to the numerical 
solution $R(t)$ (solid line) of the vortex annihilation model as shown in Fig.~2b.  
We find that $L_E (t)$ and $R(t)$ agree very closely with $L(t)$. 
(b) The vortex number density $\rho (t)$ versus $L(t)$ exhibits a power law relation with
$\rho(t) \sim L(t)^{-1.92(1)}$ consistent with the small discrepancy between 
$L(t)$ and $L_V(t)$ in (a).}
\label{Length_scales}
\end{figure}

\begin{figure}[t]
\includegraphics[angle=0,width=8cm]{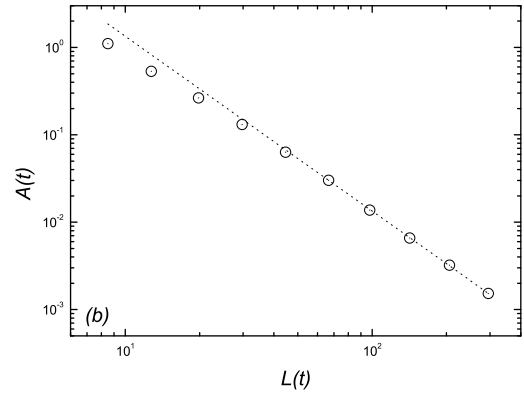}
\caption{ 
Nonequilibrium autocorrelation function $A(t)$ of the $O(2)$ order parameter 
 versus the growing length scale $L(t)$ (open circles), which exhibits a relation of 
$A(t) \sim L(t)^{\lambda}$ with $\lambda \simeq 1.99(2)$ in the long time limit.  } \label{autoco_vs_L}
\end{figure}

\end{document}